\documentclass[twocolumn,showpacs,amsmath,amssymb]{revtex4}
\input{epsf}

\usepackage{graphicx}
\usepackage{array}
\usepackage{bigstrut}
\usepackage{longtable}
\usepackage{rotating,booktabs}
\usepackage{booktabs,threeparttable}
\usepackage{bm}
\usepackage{lipsum}

\begin{document}

\title{Magic intensity trapping of the Mg lattice clock with light shift suppressed below $10^{-19}$}

\author{Fang-Fei Wu,$^{1,4}$ Yong-Bo Tang,$^{2,3,*}$~\footnotetext{*Email Address: ybtang@htu.edu.cn} Ting-Yun Shi,$^{1}$ and Li-Yan Tang$^{1,\dag}$~\footnotetext{\dag Email Address: lytang@wipm.ac.cn}}

\affiliation {$^1$ State Key Laboratory of Magnetic Resonance and
Atomic and Molecular Physics, Wuhan Institute of Physics and Mathematics, Chinese Academy of Sciences, Wuhan 430071, People's Republic of China}
\affiliation {$^2$ College of Engineering Physics, Shenzhen technology University, Shenzhen 518118, People's Republic of China}
\affiliation {$^3$ College of Physics and Materials Science, Henan Normal University, Xinxiang 453007, People's Republic of China}
\affiliation {$^4$ University of Chinese Academy of Sciences, Beijing 100049, People's Republic of China}

\date{\today}

\begin{abstract}
Progress in atomic optical clocks with total uncertainty of $10^{-18}$ or below requires a precise estimation of multipolar and higher-order effects due to atom-field interactions. Magnesium is an attractive candidate for optical lattice clocks because it is insensitive to blackbody radiation and has a large quality factor. We employ a combined method of the Dirac-Fock plus core polarization and the relativistic configuration interaction to calculate the dynamic multipolar polarizabilities and the hyperpolarizabilities of the atomic Mg clock. The lattice light shift against variation of the laser detuning and trap depth is also investigated. We find that there exists a distinctive operational magic lattice intensity of $5.33(2)E_R$ ($E_R$ is the lattice photon recoil energy) that reduces the total light shift below $1\times 10^{-19}$ over 14\% of the trap depth variation, which will pave the way for the development of a new time-frequency standard of the Mg lattice clock.
\end{abstract}

\maketitle

In the past few decades, with rapid development of laser cooling and trapping techniques, there have been tremendous advance in improving performance of atomic optical clocks. Today's leading optical clocks are based on electronic-dipole forbidden transitions in some selected atoms and singly charged ions~\cite{nicholson15a,campbell17a,brewer19a,mcgrew18a,huang16a}. Properly controlling relevant degrees of freedom in these atoms and ions has resulted in an unprecedented precision in clock transition frequencies. For example, the systematic uncertainty for the Al$^+$ clock has been reduced to $9.4\times10^{-19}$~\cite{brewer19a}, the systematic uncertainty for the Yb$^+$ clock has been reduced to
$3.2\times10^{-18}$ ~\cite{huntemann16a}, and the uncertainties for the Sr and Yb lattice clocks have been reduced to $10^{-18}$~\cite{mcgrew18a,nicholson15a}. It is strongly desirable to develop atomic clocks with even higher precision so that they could be used for performing precision measurements of fundamental physical constants~\cite{bregolin17a,yamanaka15a}, for exploring possible variation of these constants~\cite{godun14a,huntemann14a,safronova18a}, for probing a violation of the local Lorentz invariance~\cite{bars17a,shaniv18a}, and for detecting new forces beyond the standard model of particle physics~\cite{kolkowitz16a,roberts17a}.

Meticulously controlling the interaction between the target atom and external field becomes the heart of developing next-generation high-precision optical clocks. For current neutral atomic clocks, eliminating the second-order light shift in clock transition frequencies, by designing an optical lattice operating at the magic wavelength, results in the systematic uncertainty below $10^{-17}$~\cite{nicholson15a,mcgrew18a,brown17a}. However, for pursuing a higher precision, such type of elimination is insufficient
because of non-negligible contributions from the multipolar and higher-order light shifts, which are related to the electric quadrupole polarizability, the magnetic dipole polarizability, and the hyperpolarizability. Therefore, suppressing these light shifts is now a crucial issue.

The concept of magic ellipticity is proposed with the purpose of removing the fourth-order light shift that is directly related to the hyperpolarizability~\cite{katori15a,taichenachev06a}. The criterion for having such a magic ellipticity is that the signs of the differential hyperpolarizabilities under the linearly and circularly polarized lights are opposite~\cite{katori15a,ovsiannikov13a,taichenachev06a}. However, not all of atomic clocks satisfy this criterion. Recently, the magic intensity trapping is proposed to make the overall light shift of the lattice clocks insensitive to lattice-intensity variation, and also has been applied into the developing a single neutral atom qubit~\cite{zhan16a,zhan18a}. For the current lattice clocks, only Yb and Sr clocks are predicted existing an operational magic intensity~\cite{brown17a,ushijima18a}, which can suppress the total light shift down to the level of $10^{-19}$. In order to find out such a magic condition, precise knowledge of multipolar polarizabilities and hyperpolarizabilities is required.

Atomic magnesium has been proposed as one of new potential candidates for developing a time-frequency standard due to its unique properties. Compared to other neutral atomic optical clocks, the blackbody radiation shift of Mg at the room temperature is about $3.9\times 10^{-16}$~\cite{kulosa15a}, which is one order of magnitude smaller than the Sr and Yb lattice clocks~\cite{middelmann12a,sherman12a}. The quality factor $Q$ of Mg is estimated as $\sim7.1\times10^{18}$ from the transition frequency and the lifetime of the excited $^3P_0^o$ state~\cite{nistasd500}, which is one or two orders of magnitude larger than that of the Sr, Yb, and Hg lattice clocks~\cite{porsev08a,safronova18a,porsev17a}. In 2015, Kulosa {\em et al.} carried out the measurements of the $3s^2\,^1S_0\rightarrow 3s3p\,^3P_0$ transition frequency, the Zeeman and Stark frequency shifts of Mg. In their work, they also stated that lattices with a larger depth of more than 40 photon recoil energies would allow more precise spectroscopic measurements~\cite{kulosa15a}. Under such larger lattice depth, the multipolar and higher-order Stark shifts, which are determined by the multipolar polarizabilities and hyperpolarizabilities, need to be known quantitatively for further development of the Mg clock.

Because of the scarcity of theoretical data and the difficulty in experiment, none of accurate values for the multipolar polarizabilities and hyperpolarizabilities have been reported for Mg, except the one using the single-electron Fues' model-potential (FMP) approach~\cite{ovsiannikov17a}. However, the predictive ability of the single-electron FMP approach for many-electron atoms is limited. In fact, for the multipolar polarizabilities and hyperpolarizabilities of the Sr clock, the results obtained from the single-electron FMP~\cite{ovsiannikov17a} show a large discrepancy with the configuration interaction (CI) calculations~\cite{porsev18a}. As we know, for multi-electron atoms, a full-electron calculation of multipolar polarizabilities and hyperpolarizabilities is challenging, because of the electron-electron correlations among all electrons. Thus, an effective many-electron atomic theory and efficient computational technique are needed to ensure that the calculations of the multipolar polarizabilities and hyperpolarizabilities for the Mg clock are accurate and reliable.

The purpose of this paper is to study high-order external field effects on the Mg lattice clock, using the Dirac-Fock plus core-polarization (DFCP) method together with the relativistic configuration interaction (RCI) method to treat divalent atoms. This combined method adopts the mean-field approximation to handle the electron-electron interactions in the core part. The interaction between two valence electrons is added into the Dirac-Coulomb Hamiltonian directly. We apply this approach to perform relativistic calculations for dynamic multipolar polarizabilities and hyperpolarizabilities of Mg. We analyse the total light shift of the Mg clock by using the obtained multipolar polarizabilities and hyperpolarizabilities, and derive a distinctive operational magic intensity that can reduce the total light shift to a level of $10^{-19}$ or below.

\textbf{DFCP+RCI method} The basic strategy of our theoretical method is to simplify a divalent electron atom into a frozen core and two valence electrons. The detailed calculations can be divided into three steps. The first step is to carry out the Dirac-Fock (DF) calculation for the frozen core to obtain the core orbital functions~\cite{tang14a}, which are used to calculate the matrix elements of the DF potential. The second step is to solve the DFCP equation to obtain the wave functions of the monovalent electron ion, which are used to construct the configuration wave functions. In this step, the static dipole polarizability of the Mg$^{2+}$ ion~\cite{opik67a} is 0.489 a.u., which is adopted to construct a semi-empirical one-body core-polarization potential. The third step is to implement the configuration interaction calculation for a divalent electron atom. The Notre Dame basis sets~\cite{johnson88a} are used in numerical calculations. The detailed description of the DFCP+RCI method can be seen in the Ref.~\cite{wu19b}.

\textbf{Lattice light shift} For atoms trapped under a 1D optical lattice with the laser frequency $\omega$ and the linearly polarized laser field intensity $I$, the light shift for a clock transition is expressed as~\cite{ushijima18a}
%
\begin{eqnarray}
h\Delta\nu&=&\bigg[\frac{\partial\Delta\alpha^{E1}(\omega)}{\partial\nu}\delta-\Delta\alpha^{QM}(\omega)\bigg](n_z+\frac{1}{2})\sqrt{\frac{E_R}{\alpha^{E1}(\omega)}}I^{1/2} \nonumber \\
&-&\bigg[\frac{\partial\Delta\alpha^{E1}(\omega)}{\partial\nu}\delta+\frac{3}{8}\frac{E_R\Delta\gamma_0^{\ell}(\omega)}{\alpha^{E1}(\omega)}(n_z^2+n_z+\frac{1}{2})\bigg]I \nonumber \\
&+&\frac{1}{2}\Delta\gamma_0^{\ell}(\omega)\sqrt{\frac{E_R}{\alpha^{E1}(\omega)}}(n_z+\frac{1}{2})I^{3/2}-\frac{1}{4}\Delta\gamma_0^{\ell}(\omega) I^2 \,, \nonumber \\
\label{e16}
\end{eqnarray}
%
where $\delta=\nu-\nu_m$ is the frequency detuning of the lattice laser frequency $\nu$ relative to the magic frequency $\nu_m=\omega_m/2\pi$, here $\omega_m$ is determined by making the differential $E1$ polarizability of the clock transition $\Delta\alpha^{E1}(\omega)=0$, $n_z$ is the vibrational state of atoms along the $z$ axis, $E_R=h^2/(2\mathcal{M}\lambda_m^2)$ is the lattice photon recoil energy with $\mathcal{M}$ being the atomic mass. $\Delta\alpha^{QM}(\omega)=\Delta\alpha^{M1}(\omega)+\Delta\alpha^{E2}(\omega)$ is the differential $M1$-$E2$ polarizability, and $\Delta\gamma_0^{\ell}(\omega)$ represents the differential hyperpolarizability between the upper and lower energy levels associated with the clock transition.

For an initial state $|0\rangle\equiv|n_0,J_0=0\rangle$, where $n_0$ denotes all other quantum numbers, the dynamic $2^k$-pole polarizability in the atomic units (a.u.) can be written as~\cite{porsev04a}
\begin{eqnarray}
\alpha^{\lambda k}(\omega)&=&\frac{(k+1)(2k+1)}{k[(2k+1)!!]^2}(\alpha\omega)^{2k-2}\nonumber \\
&&\times\sum_n\frac{\Delta E_{n0}|\langle 0\|T_{\lambda k}\|nJ_n\rangle|^2}{\Delta E_{n0}^2-\omega^2}\,, \label{e1}
\end{eqnarray}
where $\alpha$ is the fine structure constant, $\Delta E_{n0}$ is the transition energy between the initial state $|0\rangle$ and the intermediate state $|nJ_n\rangle$, $\lambda$ distinguishes the electric ($\lambda=E$) and magnetic ($\lambda=M$) multipoles, $T_{E1}\equiv D$, $T_{M1}\equiv \mu$, and $T_{E2}\equiv Q$ are respectively the $E1$, $M1$, and $E2$ transition operators.

The dynamic hyperpolarizabilities $\gamma_0^{\ell}(\omega)$ and $\gamma_0^{c}(\omega)$ under the linearly and circularly polarized lights for the initial state $|0\rangle\equiv|n_0,J_0=0\rangle$ can be written as
\begin{eqnarray}
\gamma_0^{\ell}(\omega)=\frac{1}{9}\mathcal{T}(1,0,1,\omega,-\omega,\omega)+\frac{2}{45}\mathcal{T}(1,2,1,\omega,-\omega,\omega)\,, \\
\gamma_0^{c}(\omega)=\frac{1}{9}\mathcal{T}(1,0,1,\omega,-\omega,\omega)+\frac{1}{90}\mathcal{T}(1,2,1,\omega,-\omega,\omega)\,,
\label{e3b}
\end{eqnarray}
where $\mathcal{T}(J_a,J_b,J_c,\omega_1,\omega_2,\omega_3)$ is expanded as the followed formula~\cite{tang14b},
\begin{widetext}
\begin{eqnarray}
\mathcal{T}(J_a,J_b,J_c,\omega,-\omega,\omega)&=&4\sum_{n_an_bn_c}^{\prime}\langle 0\|D\|n_aJ_a\rangle\langle n_aJ_a\|D\|n_bJ_b\rangle \langle n_bJ_b\|D\|n_cJ_c\rangle\langle n_cJ_c\|D\|0\rangle \nonumber \\
&&\Bigg[\frac{1}{(\Delta E_{n_a0}-\omega)(\Delta E_{n_b0}-2\omega)(\Delta E_{n_c0}-\omega)}+\frac{1}{(\Delta E_{n_a0}+\omega)(\Delta E_{n_b0}+2\omega)(\Delta E_{n_c0}+\omega)} \nonumber \\
&&+\frac{4\Delta E_{n_a0}\Delta E_{n_c0}}{(\Delta E_{n_a0}+\omega)(\Delta E_{n_a0}-\omega)\Delta E_{n_b0}(\Delta E_{n_c0}+\omega)(\Delta E_{n_c0}-\omega)}\Bigg]+8(-1)^{J_a+J_c+1}\delta(J_b,J_0)\nonumber \\
&&\sum_{n_a}\frac{\Delta E_{n_a0}|\langle0\|D\|n_aJ_a\rangle|^2}{\Delta E_{n_a0}^2-\omega^2}\sum_{n_c}\frac{(3\Delta E_{n_c0}^2+\omega^2)|\langle0\|D\|n_cJ_c\rangle|^2}{(\Delta E_{n_c0}^2-\omega^2)^2}
\,,\label{e5}
\end{eqnarray}
\end{widetext}
where the prime over the summation means that the intermediate state $|n_iJ_i\rangle\equiv|n_0,J_0=0\rangle$ ($i=a,b,c$) should be excluded.

Since the term $\mathcal{T}(J_a,J_b,J_c,\omega_1,\omega_2,\omega_3)$ involves three summations over a large number of intermediate states, and the completeness of the intermediate states is vital for the reliability of the calculations, the accurate calculations of the dynamic hyperpolarizabilities are challenging. In this work, we apply the DFCP+RCI method to perform a large-scale configuration-interaction calculation to obtain the matrix elements by including sufficient configurations in an appropriate cavity to make sure the completeness of intermediate states, and use the sum-over-state approach to calculate the dynamic multipolar polarizabilities and hyperpolarizabilities of Mg by replacing the energy levels of low-lying states with the NIST values~\cite{nistasd500}, so that the main source of our uncertainty comes from the reduced matrix elements.

\textbf{Detailed comparisons for the energies, reduced matrix elements and static dipole polarizabilities}
Table~\ref{t1} presents a comparison of the energy for some of low-lying states of the Mg atom between present DFCP+RCI calculations and NIST energy~\cite{nistasd500}. All the energies are given relative to the ground state of the Mg$^{2+}$ core. The relative difference (Diff.) between present energy and the NIST energy [26] is difined as Diff.$=$(Present-NIST)/NIST$\times$100\%. The biggest difference between our results and NIST energy is 0.097\%. Table~\ref{t2} presents a comparison of some selective reduced matrix elements. Table~\ref{t3} presents the static dipole polarizabilities for the clock states of $3s^2\,^1S_0$ and $3s3p\,^3P_0^o$ of the Mg atom. For the static electric dipole polarizability, our values are 71.485 a.u. and 101.327 a.u. for the $3s^2\,^1S_0$ and $3s3p\,^3P_0$ clock states, respectively, which agree well with the results of 71.251 a.u. and 100.922 a.u. of Ref.~\cite{kulosa15a} within 0.5\%. The detailed and systematic comparison of the electric dipole polarizability for the ground state of the Mg atom is given by J. F. Babb in Ref.~\cite{babb15a}.

It is seen from Table~\ref{t2} that the difference for all the reduced matrix elements between our results and the values of Ref.~\cite{nistasd500} are within 3\% except the $3s^2\,^1S_0\rightarrow 3s4p\,^1P_1^o$ and $3s^2\,^1S_0\rightarrow 3s5p\,^1P_1^o$ transitions. However, from Table~\ref{t3}, we can see that the contributions from both of $3s^2\,^1S_0\rightarrow 3s4p\,^1P_1^o$ and $3s^2\,^1S_0\rightarrow 3s5p\,^1P_1^o$ transitions to the ground-state polarizability of the Mg clock are much smaller than the 95\% contribution from the $3s^2\,^1S_0\rightarrow3s3p\,^1P_1^o$ transition. This indicates the large difference in the both reduced matrix elements of $3s^2\,^1S_0\rightarrow 3s4p\,^1P_1^o$ and $3s^2\,^1S_0\rightarrow 3s5p\,^1P_1^o$ transitions between present work and Ref.~\cite{nistasd500} has insignificant effect on the final polarizability. Therefore, we introduce $\pm$3\% fluctuation into all the reduced matrix elements to estimate conservatively the uncertainties of multipolar polarizabilities and hyperpolaribilities for the Mg clock. This method of evaluating uncertainty has been extensively used for the Ca$^+$, Sr$^+$, and K atomic systems~\cite{tang13b,jiang16a,jiang13a}.

\begin{table}[!htbp]
\caption{\label{t1} The energies (in cm$^{-1}$) for some of the low-lying states of the Mg atom. All the energies are given relative to the ground state of the Mg$^{2+}$ core. The relative differences between present energy and the NIST energy~\cite{nistasd500} are listed as Diff.}
\begin{ruledtabular}
\begin{tabular}{l@{\extracolsep{2em}}l@{\extracolsep{2em}}l@{\extracolsep{2em}}l@{\extracolsep{2em}}l@{\extracolsep{2em}}l@{\extracolsep{2em}}l@{\extracolsep{2em}}l}
\multicolumn{1}{c}{State}&\multicolumn{1}{c}{Present}&\multicolumn{1}{c}{NIST~\cite{nistasd500}}&\multicolumn{1}{c}{Diff.}
\\ \hline
$3s^{2}$ $^{1}S_{0}$ &$-$182762.10  & $-$182938.66 & $-$0.097\%    \\
$3s4s  $ $^{1}S_{0}$ &$-$139387.35  & $-$139435.34 & $-$0.034\%   \\
$3s5s  $ $^{1}S_{0}$ &$-$130364.75  & $-$130382.46 & $-$0.014\%    \\
$3s6s  $ $^{1}S_{0}$ &$-$126743.61  & $-$126751.80 & $-$0.006\%   \\
$3p^{2}$ $^{3}P_{0}$ &$-$125134.93  & $-$125125.90 & $-$0.007\%  \\
$3s7s  $ $^{1}S_{0}$ &$-$124924.96  & $-$124929.27 & $-$0.003\%   \\
$3s8s  $ $^{1}S_{0}$ &$-$123882.68  & $-$123885.15 & $-$0.002\%  \\
$3s9s  $ $^{1}S_{0}$ &$-$123230.07  & $-$123231.56 & $-$0.001\%  \\
\hline
$3s3p  $ $^{3}P_{0}$ &$-$160969.42  & $-$161088.26 & $-$0.074\% \\
$3s4p  $ $^{3}P_{0}$ &$-$135076.16  & $-$135097.55 & $-$0.016\% \\
$3s5p  $ $^{3}P_{0}$ &$-$128681.61  & $-$128689.86 & $-$0.006\% \\
$3s6p  $ $^{3}P_{0}$ &$-$125917.64  & $-$125921.59 & $-$0.003\% \\
$3s7p  $ $^{3}P_{0}$ &$-$124459.92  & $-$124461.98 & $-$0.002\% \\
$3s8p  $ $^{3}P_{0}$ &$-$123595.49  & $-$123596.17 & $-$0.001\% \\
$3s9p  $ $^{3}P_{0}$ &$-$123040.48  & $-$123040.81 & $-$0.0003\%\\
\hline
$3s3p  $ $^{3}P_{1}$ &$-$160949.30  & $-$161068.20 & $-$0.074\%   \\
$3s3p  $ $^{1}P_{1}$ &$-$147885.07  & $-$147887.40 & $-$0.002\%    \\
$3s4p  $ $^{3}P_{1}$ &$-$135072.85  & $-$135094.26 & $-$0.016\%   \\
$3s4p  $ $^{1}P_{1}$ &$-$133590.02  & $-$133591.94 & $-$0.001\%    \\
$3s5p  $ $^{3}P_{1}$ &$-$128680.32  & $-$128688.58 & $-$0.006\%   \\
$3s5p  $ $^{1}P_{1}$ &$-$128231.04  & $-$128232.13 & $-$0.001\%    \\
$3s6p  $ $^{3}P_{1}$ &$-$125917.00  & $-$125920.95 & $-$0.003\%    \\
\hline
$3s3p  $ $^{3}P_{2}$ & $-$160908.96 & $-$161027.49 & $-$0.074\%  \\
$3s4p  $ $^{3}P_{2}$ & $-$135066.16 & $-$135087.51 & $-$0.016\%  \\
$3s5p  $ $^{3}P_{2}$ & $-$128677.70 & $-$128685.94 & $-$0.006\%  \\
$3s4f  $ $^{3}F_{2}$ & $-$128261.89 & $-$128262.01 & $-$0.0001\% \\
$3s6p  $ $^{3}P_{2}$ & $-$125915.69 & $-$125919.65 & $-$0.003\%  \\
$3s5f  $ $^{3}F_{2}$ & $-$125734.49 & $-$125734.44 & $-$0.00004\% \\
$3s7p  $ $^{3}P_{2}$ & $-$124458.81 & $-$124460.91 & $-$0.002\%  \\
\hline
$3s4s  $ $^{3}S_{1}$ &$-$141729.94 & $-$141741.27& $-$0.008\%      \\
$3s3d  $ $^{3}D_{1}$ &$-$134977.14 & $-$134981.61& $-$0.003\%      \\
$3s5s  $ $^{3}S_{1}$ &$-$131062.36 & $-$131066.14& $-$0.003\%      \\
$3s4d  $ $^{3}D_{1}$ &$-$128744.92 & $-$128746.34& $-$0.001\%     \\
$3s6s  $ $^{3}S_{1}$ &$-$127045.31 & $-$127046.87& $-$0.001\%      \\
$3s5d  $ $^{3}D_{1}$ &$-$125969.90 & $-$125970.40& $-$0.0004\%     \\
\hline
$3s3d  $ $^{1}D_{2}$ &$-$136501.02 & $-$136535.61& $-$0.025\% \\
$3s3d  $ $^{3}D_{2}$ &$-$134977.05 & $-$134981.64& $-$0.003\% \\
$3s4d  $ $^{1}D_{2}$ &$-$129788.49 & $-$129804.03& $-$0.012\% \\
$3s4d  $ $^{3}D_{2}$ &$-$128744.91 & $-$128746.38& $-$0.001\% \\
$3s5d  $ $^{1}D_{2}$ &$-$126623.16 & $-$126630.29& $-$0.006\% \\
$3s5d  $ $^{3}D_{2}$ &$-$125969.90 & $-$125970.42& $-$0.0004\% \\
\end{tabular}
\end{ruledtabular}
\end{table}

\begin{table}[!htbp]
\caption{\label{t2} The reduced matrix elements (in a.u.) for the Mg clock. The last column is the difference between present values and the results of Ref.~\cite{nistasd500}.}
\begin{ruledtabular}
\begin{tabular}{llll}
\multicolumn{1}{c}{Transition}&\multicolumn{1}{c}{Present}&\multicolumn{1}{c}{Ref.~\cite{nistasd500}}&\multicolumn{1}{c}{Diff.}\\ \hline
$3s^2\,^1S_0\rightarrow 3s3p\,^1P_1^o$&4.037 &4.112&$-$1.82\%\\
$3s^2\,^1S_0\rightarrow 3s4p\,^1P_1^o$&0.835 &0.868&$-$3.80\%\\
$3s^2\,^1S_0\rightarrow 3s5p\,^1P_1^o$&0.365 &0.380&$-$3.95\%\\
$3s3p\,^3P_0^o\rightarrow 3s4s\,^3S_1$&1.539 &1.516&1.52\%\\
$3s3p\,^3P_0^o\rightarrow 3s3d\,^3D_1$&2.811 &2.735&2.78\%\\
$3s3p\,^3P_0^o\rightarrow 3s5s\,^3S_1$&0.419 &0.432&$-$3.00\%\\
$3s3p\,^3P_0^o\rightarrow 3s4d\,^3D_1$&1.130 &1.164&$-$2.92\%\\
$3s3p\,^3P_0^o\rightarrow 3s5d\,^3D_1$&0.665 &0.670&$-$0.75\%\\
$3s3p\,^3P_0^o\rightarrow 3p^2\,^3P_1$&2.381 &2.405&$-$1.00\%\\
\end{tabular}
\end{ruledtabular}
\end{table}
\begin{table}[!htbp]
\caption{\label{t3} Contributions (Contr.) to the static dipole polarizabilities (in a.u.) for the $3s^2\,^1S_0$ and $3s3p\,^3P_0^o$ states of the Mg clock.}
\begin{ruledtabular}
\begin{tabular}{llllll}
\multicolumn{3}{c}{$3s^2\,^1S_0$ state} &\multicolumn{3}{c}{$3s3p\,^3P_0^o$ state}\\
\cline{1-3}\cline{4-6}
\multicolumn{1}{c}{Contr.} &\multicolumn{1}{c}{Present}&\multicolumn{1}{c}{Ref.~\cite{kulosa15a}}&\multicolumn{1}{c}{Contr.} &\multicolumn{1}{c}{Present}&\multicolumn{1}{c}{Ref.~\cite{kulosa15a}}\\ \hline
$3s3p\,^1P_1^o$ &68.021 &   &$3s4s\,^3S_1$ &17.911  & \\
$3s4p\,^1P_1^o$ &2.069  &   &$3s3d\,^3D_1$ &44.300   &  \\
$3s5p\,^1P_1^o$ &0.357  &   &$3s5s\,^3S_1$ &0.854   & \\
$3s6p\,^1P_1^o$ &0.118  &   &$3s4d\,^3D_1$ &5.778    & \\
$3s7p\,^1P_1^o$ &0.053  &   &$3s5d\,^3D_1$ &1.843 & \\
$3s8p\,^1P_1^o$ &0.029  &   &$3p^2\,^3P_1$ &23.056 &\\
Remainder                              &0.349  &    &Remainder                               &7.096  &\\
Valence                                &70.996&    &Valance                                 &100.838  &\\
Core                                   &0.489   &  &Core                                    &0.489    & \\
Total                                  &71.485  &71.251 &Total                                   &101.327  &100.922 \\
\end{tabular}
\end{ruledtabular}
\end{table}
\begin{table}[ht]
\caption{\label{t4} Dynamic multipolar polarizabilities and hyperpolarizabilities (in a.u.) at the 468.46(21)~nm magic wavelength for the $3s^2\,^1S_0$ and $3s3p\,^3P_0^o$ clock states. $\alpha^{QM}(\omega)=\alpha^{M1}(\omega)+\alpha^{E2}(\omega)$. The last column lists the differential between the two clock states. Numbers in parentheses are computational uncertainties. Numbers in square brackets represent the power of 10.}
\begin{ruledtabular}
\begin{tabular}{lccc}
&\multicolumn{1}{c}{$3s^2\,^1S_0$} &\multicolumn{1}{c}{$3s3p\,^3P_0^o$} &\multicolumn{1}{c}{$\Delta(3s^2\,^1S_0-3s3p\,^3P_0^o)$}  \\ \hline
$\alpha^{M1}(\omega)$     &9.93(60)[-12]  &$-$1.72(10)[-7]   &$-$1.72(10)[-7]  \\
$\alpha^{E2}(\omega)$     &4.25(26)[-5]   &9.90(59)[-5]    &5.65(64)[-5]  \\
$\gamma_0^{\ell}(\omega)$ &6.57(81)[6]    &1.74(22)[8]     &1.67(22)[8] \\
                          &                 &                  &8.28[7]~\cite{ovsiannikov17a} \\
$\gamma_0^{c}(\omega)$    &5.55(69)[6]      &3.95(50)[7]       &3.39(50)[7] \\
$\alpha^{QM}(\omega)$     &4.25(26)[-5]   &9.88(59)[-5]    &5.63(64)[-5] \\
                          &                 &                  &2.92[-5]~\cite{ovsiannikov17a} \\
\end{tabular}
\end{ruledtabular}
\end{table}

\textbf{Multipolar polarizabilities and hyperpolarizabilities under the linearly polarized light}
The dynamic multipolar polarizabilities and hyperpolarizabilities at the 468.46(21)~nm magic wavelength~\cite{kulosa15a} for the $3s^2\,^1S_0$ and $3s3p\,^3P_0^o$ clock states are listed in Table~\ref{t4}. For the $3s^2\,^1S_0$ state, the dynamic $M1$ polarizability is seven orders of magnitude smaller than the dynamic $E2$ polarizability, and five orders of magnitude smaller than the $M1$ polarizability of the $3s3p\,^3P_0^o$ state. Thus, the contribution from $\alpha^{M1}(\omega)$ of $3s^2\,^1S_0$ to the differential $M1$ polarizability $\Delta\alpha^{M1}(\omega)$ is negligible. Compared to the $E2$ polarizability, $\Delta\alpha^{M1}(\omega)$ is two orders of magnitude smaller than $\Delta\alpha^{E2}(\omega)$, which causes the final value of $\Delta\alpha^{QM}(\omega)=5.63(64)\times10^{-5}$~a.u. to be determined largely by $\Delta\alpha^{E2}(\omega)$. The differential hyperpolarizability $\Delta\gamma_0^{\ell}(\omega)$ is $1.67(22)\times10^{8}$~a.u., which comes mainly from the dynamic hyperpolarizability of the $3s3p\,^3P_0^o$ state. The values from the single-electron FMP approach for $\Delta\alpha^{QM}(\omega)$ and $\Delta\gamma_0^{\ell}(\omega)$ are $2.92\times 10^{-5}$~a.u. and $8.28\times 10^7$~a.u.~\cite{ovsiannikov17a}, respectively. It is seen that our values for $\Delta\alpha^{QM}(\omega)$ and $\Delta\gamma_0^{\ell}(\omega)$ are about twice as large as the results of the single-electron FMP approach~\cite{ovsiannikov17a}.

\begin{figure}
\includegraphics[width=0.49\textwidth]{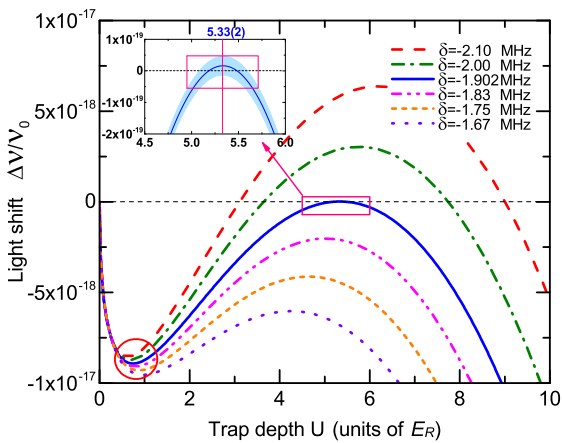}
\caption{\label{f3}(Color online) Lattice light shift of the Mg clock for different detuning $\delta$ as the trap depth $U$ increased. The circle area indicates the effect of the high-order Stark shift starts to be noticeable. The inset shows the position of the operational magic intensity of $U=5.33(2)E_R$ for the detuning $\delta=-1.902(1)$~MHz, where the light shift is zero. The blue shaded area indicates an uncertainty given by 0.001~MHz deviation for the detuning of the blue line. The light shift variation in the rectangular area is less than $1\times10^{-19}$ over the trap depth $4.95E_R<U<5.71E_R$. }
\end{figure}

\textbf{Magic ellipticity and operational magic intensity}
In order to analysis the lattice light shift of the Mg clock using the Eq.~(\ref{e16}), the Planck constant $h=1$ is adopted, the units of the qualities $\alpha^{E1}(\omega)$, $\Delta \alpha^{E1}(\omega)$, $\Delta \alpha^{QM}(\omega)$ are transformed from the atomic units to Hz/(kW/cm$^2$), and the units of $\Delta \gamma_0^{\ell}(\omega)$, $E_R$ and $I$ are $\mu$Hz/(kW/cm$^2$)$^2$, Hz and kW/cm$^2$, respectively. Then we obtain the photon recoil energy $E_R=37.9$~kHz, at the magic wavelength of 468.46(21) nm, the dipole polarizability is $\alpha_1(\omega)=112(7)$~a.u. $=21(2)$ kHz/(kW/cm$^2$), the rate of change for the differential dipole polarizability is $\partial\Delta\alpha^{E1}(\omega)/\partial\nu=2.56(54)\times 10^{-9}$ (kW/cm$^2$)$^{-1}$, the differential $M1$-$E2$ polarizability is $\Delta \alpha^{QM}(\omega)=10.6(1.2)$ mHz/(kW/cm$^2$), and the differential hyperpolarizability is $\Delta \gamma_0^{\ell}(\omega)=892(118)$~$\mu$Hz/(kW/cm$^2$)$^2$. Substituting these values into the Eq.~(\ref{e16}), we can evaluate the light shift as the laser detuning $\delta$ and trap depth $U\approx\alpha^{E1}(\omega)I$ changed by assuming that all the Mg atoms are trapped in the $n_z=0$ vibrational state, as seen in Fig.~\ref{f3}. When $\delta$ changes from the red-detuning to the blue-detuning at the range of $U< 1E_R$ area, since the influence of the higher-order Stark shift appears to be noticeable at smaller laser intensity, which indicates that the elimination of the higher-order light shift connected to the hyperpolarizabilities becomes important for the development of the Mg lattice clock.

The concept of magic ellipticity was originally proposed to eliminate higher-order effect of the hyperpolarizability~\cite{katori15a,taichenachev06a}, which exists only when the signs of the differential hyperpolarizabilities for the linearly and circularly polarized lights are opposite. For the Mg clock, however, the differential hyperpolarizabilities at the magic wavelength are $3.39(50)\times10^{7}$~a.u. and $1.67(22)\times10^{8}$~a.u., under
the circularly and linearly polarized light, respectively, it illustrates that there does not exist a magic ellipticity for a direct cancellation of higher-order Stark shift~\cite{katori15a,taichenachev06a}.

In order to locate an operational magic intensity at which the total shift is insensitive to the trap depth, for each curve in Fig.~\ref{f3}, we use the condition $(\partial \Delta\nu/\partial U)|_{U=U_{\rm op}}=0$ to determine the position of the operational magic intensity. For the range marked by a pink rectangle in Fig.~\ref{f3}, there exists a distinctive operational condition $U_{\rm op}=5.33(2)E_R$ with $\delta_{\rm op}=-1.902(1)$~MHz that makes the variation of total light shift below $1.0\times10^{-19}$ over the trap depth range $4.95E_R<U<5.71E_R$. For the blue line, the blue shaded area is given by 0.001~MHz deviation from the central detuning $\delta=-1.902$~MHz. The present operational trap depth and its allowable intensity range are feasible to be implemented experimentally~\cite{kulosa15a}. Therefore, the operational condition $U_{\rm op}=5.33(2)E_R$ with $\delta_{\rm op}=-1.902(1)$~MHz predicted in this work can provide a reference for developing the Mg optical lattice clock at the level of $10^{-19}$.

In summary, we have applied the DFCP+RCI method to calculate the dynamic magnetic dipole and electric quadrupole polarizabilities and the hyperpolarizabilities at the magic wavelength for the $3s^2\,^1S_0$ and $3s3p\,^3P_0^o$ clock states of the Mg atom. The present differential multipolar polarizability and differential hyperpolarizability under the linearly polarized light are $5.63(64)\times10^{-5}$~a.u. and $1.67(22)\times10^8$~a.u., respectively, which are about twice as much as the results of the single-electron FMP approach~\cite{ovsiannikov17a}. Furthermore, we have illustrated that there is no operational magic ellipticity to cancel the forth-order Stark shift. Finally, we have predicted the existence of the distinctive operational condition $U_{\rm op}=5.33(2)E_R$ with $\delta_{\rm op}=-1.902(1)$~MHz that reduces the total light shift below $1\times 10^{-19}$ over the trap depth range $4.95E_R<U<5.71E_R$. Our results will provide important support for the realization of the Mg lattice clock with the fractional uncertainty at $10^{-19}$.

\begin{acknowledgments}
We thank Baolong L\"{u}, K. L. Gao, and Zhuanxian Xiong for useful comments, thank S. G. Porsev and M. S. Safronova for their communications, and thank Z.-C. Yan and Y. M. Yu for their careful reading. This work was supported by the National Key Research and Development Program of China under Grant No. 2017YFA0304402, by the Strategic Priority Research Program of the Chinese Academy of Sciences under Grant No. XDB21030300, by the National Natural Science Foundation of China under Grant No. 11774386, and by the Hubei Province Science Fund for Distinguished Young Scholars No. 2019CFA058. Y.-B. Tang was supported by the National Natural Science Foundation of China under Grant No. 11504094.
\end{acknowledgments}


\begin{thebibliography}{40}
\expandafter\ifx\csname natexlab\endcsname\relax\def\natexlab#1{#1}\fi
\expandafter\ifx\csname bibnamefont\endcsname\relax
  \def\bibnamefont#1{#1}\fi
\expandafter\ifx\csname bibfnamefont\endcsname\relax
  \def\bibfnamefont#1{#1}\fi
\expandafter\ifx\csname citenamefont\endcsname\relax
  \def\citenamefont#1{#1}\fi
\expandafter\ifx\csname url\endcsname\relax
  \def\url#1{\texttt{#1}}\fi
\expandafter\ifx\csname urlprefix\endcsname\relax\def\urlprefix{URL }\fi
\providecommand{\bibinfo}[2]{#2}
\providecommand{\eprint}[2][]{\url{#2}}

\bibitem[{\citenamefont{Nicholson et~al.}(2015)\citenamefont{Nicholson,
  Campbell, Hutson, Marti, Bloom, McNally, Zhang, Barrett, Safronova, Strouse
  et~al.}}]{nicholson15a}
\bibinfo{author}{\bibfnamefont{T.}~\bibnamefont{Nicholson}},
  \bibinfo{author}{\bibfnamefont{S.}~\bibnamefont{Campbell}},
  \bibinfo{author}{\bibfnamefont{R.}~\bibnamefont{Hutson}},
  \bibinfo{author}{\bibfnamefont{G.}~\bibnamefont{Marti}},
  \bibinfo{author}{\bibfnamefont{B.}~\bibnamefont{Bloom}},
  \bibinfo{author}{\bibfnamefont{R.}~\bibnamefont{McNally}},
  \bibinfo{author}{\bibfnamefont{W.}~\bibnamefont{Zhang}},
  \bibinfo{author}{\bibfnamefont{M.}~\bibnamefont{Barrett}},
  \bibinfo{author}{\bibfnamefont{M.}~\bibnamefont{Safronova}},
  \bibinfo{author}{\bibfnamefont{G.}~\bibnamefont{Strouse}},
  \bibnamefont{et~al.}, \bibinfo{journal}{Nat. Commun.}
  \textbf{\bibinfo{volume}{6}}, \bibinfo{pages}{6896} (\bibinfo{year}{2015}).

\bibitem[{\citenamefont{Campbell et~al.}(2017)\citenamefont{Campbell, Hutson,
  Marti, Goban, Darkwah~Oppong, McNally, Sonderhouse, Robinson, Zhang, Bloom
  et~al.}}]{campbell17a}
\bibinfo{author}{\bibfnamefont{S.~L.} \bibnamefont{Campbell}},
  \bibinfo{author}{\bibfnamefont{R.~B.} \bibnamefont{Hutson}},
  \bibinfo{author}{\bibfnamefont{G.~E.} \bibnamefont{Marti}},
  \bibinfo{author}{\bibfnamefont{A.}~\bibnamefont{Goban}},
  \bibinfo{author}{\bibfnamefont{N.}~\bibnamefont{Darkwah~Oppong}},
  \bibinfo{author}{\bibfnamefont{R.~L.} \bibnamefont{McNally}},
  \bibinfo{author}{\bibfnamefont{L.}~\bibnamefont{Sonderhouse}},
  \bibinfo{author}{\bibfnamefont{J.~M.} \bibnamefont{Robinson}},
  \bibinfo{author}{\bibfnamefont{W.}~\bibnamefont{Zhang}},
  \bibinfo{author}{\bibfnamefont{B.~J.} \bibnamefont{Bloom}},
  \bibnamefont{et~al.}, \bibinfo{journal}{Science}
  \textbf{\bibinfo{volume}{358}}, \bibinfo{pages}{90} (\bibinfo{year}{2017}).

\bibitem[{\citenamefont{Brewer et~al.}(2019)\citenamefont{Brewer, Chen, Hankin,
  Clements, Chou, Wineland, Hume, and Leibrandt}}]{brewer19a}
\bibinfo{author}{\bibfnamefont{S.~M.} \bibnamefont{Brewer}},
  \bibinfo{author}{\bibfnamefont{J.-S.} \bibnamefont{Chen}},
  \bibinfo{author}{\bibfnamefont{A.~M.} \bibnamefont{Hankin}},
  \bibinfo{author}{\bibfnamefont{E.~R.} \bibnamefont{Clements}},
  \bibinfo{author}{\bibfnamefont{C.~W.} \bibnamefont{Chou}},
  \bibinfo{author}{\bibfnamefont{D.~J.} \bibnamefont{Wineland}},
  \bibinfo{author}{\bibfnamefont{D.~B.} \bibnamefont{Hume}}, \bibnamefont{and}
  \bibinfo{author}{\bibfnamefont{D.~R.} \bibnamefont{Leibrandt}},
  \bibinfo{journal}{Phys. Rev. Lett.} \textbf{\bibinfo{volume}{123}},
  \bibinfo{pages}{033201} (\bibinfo{year}{2019}).

\bibitem[{\citenamefont{{McGrew} et~al.}(2018)\citenamefont{{McGrew}, {Zhang},
  {Fasano}, {Sch$\ddot{a}$ffer}, {Beloy}, {Nicolodi}, {Brown}, {Hinkley},
  {Milani}, {Schioppo} et~al.}}]{mcgrew18a}
\bibinfo{author}{\bibfnamefont{W.~F.} \bibnamefont{{McGrew}}},
  \bibinfo{author}{\bibfnamefont{X.}~\bibnamefont{{Zhang}}},
  \bibinfo{author}{\bibfnamefont{R.~J.} \bibnamefont{{Fasano}}},
  \bibinfo{author}{\bibfnamefont{S.~A.} \bibnamefont{{Sch$\ddot{a}$ffer}}},
  \bibinfo{author}{\bibfnamefont{K.}~\bibnamefont{{Beloy}}},
  \bibinfo{author}{\bibfnamefont{D.}~\bibnamefont{{Nicolodi}}},
  \bibinfo{author}{\bibfnamefont{R.~C.} \bibnamefont{{Brown}}},
  \bibinfo{author}{\bibfnamefont{N.}~\bibnamefont{{Hinkley}}},
  \bibinfo{author}{\bibfnamefont{G.}~\bibnamefont{{Milani}}},
  \bibinfo{author}{\bibfnamefont{M.}~\bibnamefont{{Schioppo}}},
  \bibnamefont{et~al.}, \bibinfo{journal}{Nature}
  \textbf{\bibinfo{volume}{564}}, \bibinfo{pages}{87} (\bibinfo{year}{2018}).

\bibitem[{\citenamefont{Huang et~al.}(2016)\citenamefont{Huang, Guan, Liu,
  Bian, Ma, Liang, Li, and Gao}}]{huang16a}
\bibinfo{author}{\bibfnamefont{Y.}~\bibnamefont{Huang}},
  \bibinfo{author}{\bibfnamefont{H.}~\bibnamefont{Guan}},
  \bibinfo{author}{\bibfnamefont{P.}~\bibnamefont{Liu}},
  \bibinfo{author}{\bibfnamefont{W.}~\bibnamefont{Bian}},
  \bibinfo{author}{\bibfnamefont{L.}~\bibnamefont{Ma}},
  \bibinfo{author}{\bibfnamefont{K.}~\bibnamefont{Liang}},
  \bibinfo{author}{\bibfnamefont{T.}~\bibnamefont{Li}}, \bibnamefont{and}
  \bibinfo{author}{\bibfnamefont{K.}~\bibnamefont{Gao}},
  \bibinfo{journal}{Phys. Rev. Lett.} \textbf{\bibinfo{volume}{116}},
  \bibinfo{pages}{013001} (\bibinfo{year}{2016}).

\bibitem[{\citenamefont{Huntemann et~al.}(2016)\citenamefont{Huntemann, Sanner,
  Lipphardt, Tamm, and Peik}}]{huntemann16a}
\bibinfo{author}{\bibfnamefont{N.}~\bibnamefont{Huntemann}},
  \bibinfo{author}{\bibfnamefont{C.}~\bibnamefont{Sanner}},
  \bibinfo{author}{\bibfnamefont{B.}~\bibnamefont{Lipphardt}},
  \bibinfo{author}{\bibfnamefont{C.}~\bibnamefont{Tamm}}, \bibnamefont{and}
  \bibinfo{author}{\bibfnamefont{E.}~\bibnamefont{Peik}},
  \bibinfo{journal}{Phys. Rev. Lett.} \textbf{\bibinfo{volume}{116}},
  \bibinfo{pages}{063001} (\bibinfo{year}{2016}).

\bibitem[{\citenamefont{Bregolin et~al.}(2017)\citenamefont{Bregolin, Milani,
  Pizzocaro, Rauf, Thoumany, Levi, and Calonico}}]{bregolin17a}
\bibinfo{author}{\bibfnamefont{F.}~\bibnamefont{Bregolin}},
  \bibinfo{author}{\bibfnamefont{G.}~\bibnamefont{Milani}},
  \bibinfo{author}{\bibfnamefont{M.}~\bibnamefont{Pizzocaro}},
  \bibinfo{author}{\bibfnamefont{B.}~\bibnamefont{Rauf}},
  \bibinfo{author}{\bibfnamefont{P.}~\bibnamefont{Thoumany}},
  \bibinfo{author}{\bibfnamefont{F.}~\bibnamefont{Levi}}, \bibnamefont{and}
  \bibinfo{author}{\bibfnamefont{D.}~\bibnamefont{Calonico}},
  \bibinfo{journal}{J. Phys. Conf. Ser.} \textbf{\bibinfo{volume}{841}},
  \bibinfo{pages}{012015} (\bibinfo{year}{2017}).

\bibitem[{\citenamefont{Yamanaka et~al.}(2015)\citenamefont{Yamanaka, Ohmae,
  Ushijima, Takamoto, and Katori}}]{yamanaka15a}
\bibinfo{author}{\bibfnamefont{K.}~\bibnamefont{Yamanaka}},
  \bibinfo{author}{\bibfnamefont{N.}~\bibnamefont{Ohmae}},
  \bibinfo{author}{\bibfnamefont{I.}~\bibnamefont{Ushijima}},
  \bibinfo{author}{\bibfnamefont{M.}~\bibnamefont{Takamoto}}, \bibnamefont{and}
  \bibinfo{author}{\bibfnamefont{H.}~\bibnamefont{Katori}},
  \bibinfo{journal}{Phys. Rev. Lett.} \textbf{\bibinfo{volume}{114}},
  \bibinfo{pages}{230801} (\bibinfo{year}{2015}).

\bibitem[{\citenamefont{Godun et~al.}(2014)\citenamefont{Godun, Nisbet-Jones,
  Jones, King, Johnson, Margolis, Szymaniec, Lea, Bongs, and Gill}}]{godun14a}
\bibinfo{author}{\bibfnamefont{R.~M.} \bibnamefont{Godun}},
  \bibinfo{author}{\bibfnamefont{P.~B.~R.} \bibnamefont{Nisbet-Jones}},
  \bibinfo{author}{\bibfnamefont{J.~M.} \bibnamefont{Jones}},
  \bibinfo{author}{\bibfnamefont{S.~A.} \bibnamefont{King}},
  \bibinfo{author}{\bibfnamefont{L.~A.~M.} \bibnamefont{Johnson}},
  \bibinfo{author}{\bibfnamefont{H.~S.} \bibnamefont{Margolis}},
  \bibinfo{author}{\bibfnamefont{K.}~\bibnamefont{Szymaniec}},
  \bibinfo{author}{\bibfnamefont{S.~N.} \bibnamefont{Lea}},
  \bibinfo{author}{\bibfnamefont{K.}~\bibnamefont{Bongs}}, \bibnamefont{and}
  \bibinfo{author}{\bibfnamefont{P.}~\bibnamefont{Gill}},
  \bibinfo{journal}{Phys. Rev. Lett.} \textbf{\bibinfo{volume}{113}},
  \bibinfo{pages}{210801} (\bibinfo{year}{2014}).

\bibitem[{\citenamefont{Huntemann et~al.}(2014)\citenamefont{Huntemann,
  Lipphardt, Tamm, Gerginov, Weyers, and Peik}}]{huntemann14a}
\bibinfo{author}{\bibfnamefont{N.}~\bibnamefont{Huntemann}},
  \bibinfo{author}{\bibfnamefont{B.}~\bibnamefont{Lipphardt}},
  \bibinfo{author}{\bibfnamefont{C.}~\bibnamefont{Tamm}},
  \bibinfo{author}{\bibfnamefont{V.}~\bibnamefont{Gerginov}},
  \bibinfo{author}{\bibfnamefont{S.}~\bibnamefont{Weyers}}, \bibnamefont{and}
  \bibinfo{author}{\bibfnamefont{E.}~\bibnamefont{Peik}},
  \bibinfo{journal}{Phys. Rev. Lett.} \textbf{\bibinfo{volume}{113}},
  \bibinfo{pages}{210802} (\bibinfo{year}{2014}).

\bibitem[{\citenamefont{Safronova et~al.}(2018)\citenamefont{Safronova, Porsev,
  Sanner, and Ye}}]{safronova18a}
\bibinfo{author}{\bibfnamefont{M.~S.} \bibnamefont{Safronova}},
  \bibinfo{author}{\bibfnamefont{S.~G.} \bibnamefont{Porsev}},
  \bibinfo{author}{\bibfnamefont{C.}~\bibnamefont{Sanner}}, \bibnamefont{and}
  \bibinfo{author}{\bibfnamefont{J.}~\bibnamefont{Ye}}, \bibinfo{journal}{Phys.
  Rev. Lett.} \textbf{\bibinfo{volume}{120}}, \bibinfo{pages}{173001}
  (\bibinfo{year}{2018}).

\bibitem[{\citenamefont{Pihan-Le~Bars et~al.}(2017)\citenamefont{Pihan-Le~Bars,
  Guerlin, Lasseri, Ebran, Bailey, Bize, Khan, and Wolf}}]{bars17a}
\bibinfo{author}{\bibfnamefont{H.}~\bibnamefont{Pihan-Le~Bars}},
  \bibinfo{author}{\bibfnamefont{C.}~\bibnamefont{Guerlin}},
  \bibinfo{author}{\bibfnamefont{R.~D.} \bibnamefont{Lasseri}},
  \bibinfo{author}{\bibfnamefont{J.~P.} \bibnamefont{Ebran}},
  \bibinfo{author}{\bibfnamefont{Q.~G.} \bibnamefont{Bailey}},
  \bibinfo{author}{\bibfnamefont{S.}~\bibnamefont{Bize}},
  \bibinfo{author}{\bibfnamefont{E.}~\bibnamefont{Khan}}, \bibnamefont{and}
  \bibinfo{author}{\bibfnamefont{P.}~\bibnamefont{Wolf}},
  \bibinfo{journal}{Phys. Rev. D} \textbf{\bibinfo{volume}{95}},
  \bibinfo{pages}{075026} (\bibinfo{year}{2017}).

\bibitem[{\citenamefont{Shaniv et~al.}(2018)\citenamefont{Shaniv, Ozeri,
  Safronova, Porsev, Dzuba, Flambaum, and H\"affner}}]{shaniv18a}
\bibinfo{author}{\bibfnamefont{R.}~\bibnamefont{Shaniv}},
  \bibinfo{author}{\bibfnamefont{R.}~\bibnamefont{Ozeri}},
  \bibinfo{author}{\bibfnamefont{M.~S.} \bibnamefont{Safronova}},
  \bibinfo{author}{\bibfnamefont{S.~G.} \bibnamefont{Porsev}},
  \bibinfo{author}{\bibfnamefont{V.~A.} \bibnamefont{Dzuba}},
  \bibinfo{author}{\bibfnamefont{V.~V.} \bibnamefont{Flambaum}},
  \bibnamefont{and}
  \bibinfo{author}{\bibfnamefont{H.}~\bibnamefont{H\"affner}},
  \bibinfo{journal}{Phys. Rev. Lett.} \textbf{\bibinfo{volume}{120}},
  \bibinfo{pages}{103202} (\bibinfo{year}{2018}).

\bibitem[{\citenamefont{Kolkowitz et~al.}(2016)\citenamefont{Kolkowitz,
  Pikovski, Langellier, Lukin, Walsworth, and Ye}}]{kolkowitz16a}
\bibinfo{author}{\bibfnamefont{S.}~\bibnamefont{Kolkowitz}},
  \bibinfo{author}{\bibfnamefont{I.}~\bibnamefont{Pikovski}},
  \bibinfo{author}{\bibfnamefont{N.}~\bibnamefont{Langellier}},
  \bibinfo{author}{\bibfnamefont{M.~D.} \bibnamefont{Lukin}},
  \bibinfo{author}{\bibfnamefont{R.~L.} \bibnamefont{Walsworth}},
  \bibnamefont{and} \bibinfo{author}{\bibfnamefont{J.}~\bibnamefont{Ye}},
  \bibinfo{journal}{Phys. Rev. D} \textbf{\bibinfo{volume}{94}},
  \bibinfo{pages}{124043} (\bibinfo{year}{2016}).

\bibitem[{\citenamefont{Roberts et~al.}(2017)\citenamefont{Roberts, Blewitt,
  Dailey, Murphy, Pospelov, Rollings, Sherman, Williams, and
  Derevianko}}]{roberts17a}
\bibinfo{author}{\bibfnamefont{B.~M.} \bibnamefont{Roberts}},
  \bibinfo{author}{\bibfnamefont{G.}~\bibnamefont{Blewitt}},
  \bibinfo{author}{\bibfnamefont{C.}~\bibnamefont{Dailey}},
  \bibinfo{author}{\bibfnamefont{M.}~\bibnamefont{Murphy}},
  \bibinfo{author}{\bibfnamefont{M.}~\bibnamefont{Pospelov}},
  \bibinfo{author}{\bibfnamefont{A.}~\bibnamefont{Rollings}},
  \bibinfo{author}{\bibfnamefont{J.}~\bibnamefont{Sherman}},
  \bibinfo{author}{\bibfnamefont{W.}~\bibnamefont{Williams}}, \bibnamefont{and}
  \bibinfo{author}{\bibfnamefont{A.}~\bibnamefont{Derevianko}},
  \bibinfo{journal}{Nat. Commun.} \textbf{\bibinfo{volume}{8}},
  \bibinfo{pages}{1195} (\bibinfo{year}{2017}).

\bibitem[{\citenamefont{Brown et~al.}(2017)\citenamefont{Brown, Phillips,
  Beloy, McGrew, Schioppo, Fasano, Milani, Zhang, Hinkley, Leopardi
  et~al.}}]{brown17a}
\bibinfo{author}{\bibfnamefont{R.~C.} \bibnamefont{Brown}},
  \bibinfo{author}{\bibfnamefont{N.~B.} \bibnamefont{Phillips}},
  \bibinfo{author}{\bibfnamefont{K.}~\bibnamefont{Beloy}},
  \bibinfo{author}{\bibfnamefont{W.~F.} \bibnamefont{McGrew}},
  \bibinfo{author}{\bibfnamefont{M.}~\bibnamefont{Schioppo}},
  \bibinfo{author}{\bibfnamefont{R.~J.} \bibnamefont{Fasano}},
  \bibinfo{author}{\bibfnamefont{G.}~\bibnamefont{Milani}},
  \bibinfo{author}{\bibfnamefont{X.}~\bibnamefont{Zhang}},
  \bibinfo{author}{\bibfnamefont{N.}~\bibnamefont{Hinkley}},
  \bibinfo{author}{\bibfnamefont{H.}~\bibnamefont{Leopardi}},
  \bibnamefont{et~al.}, \bibinfo{journal}{Phys. Rev. Lett.}
  \textbf{\bibinfo{volume}{119}}, \bibinfo{pages}{253001}
  (\bibinfo{year}{2017}).

\bibitem[{\citenamefont{Katori et~al.}(2015)\citenamefont{Katori, Ovsiannikov,
  Marmo, and Palchikov}}]{katori15a}
\bibinfo{author}{\bibfnamefont{H.}~\bibnamefont{Katori}},
  \bibinfo{author}{\bibfnamefont{V.~D.} \bibnamefont{Ovsiannikov}},
  \bibinfo{author}{\bibfnamefont{S.~I.} \bibnamefont{Marmo}}, \bibnamefont{and}
  \bibinfo{author}{\bibfnamefont{V.~G.} \bibnamefont{Palchikov}},
  \bibinfo{journal}{Phys. Rev. A} \textbf{\bibinfo{volume}{91}},
  \bibinfo{pages}{052503} (\bibinfo{year}{2015}).

\bibitem[{\citenamefont{Taichenachev et~al.}(2006)\citenamefont{Taichenachev,
  Yudin, Ovsiannikov, and {Pal'chikov}}}]{taichenachev06a}
\bibinfo{author}{\bibfnamefont{A.~V.} \bibnamefont{Taichenachev}},
  \bibinfo{author}{\bibfnamefont{V.~I.} \bibnamefont{Yudin}},
  \bibinfo{author}{\bibfnamefont{V.~D.} \bibnamefont{Ovsiannikov}},
  \bibnamefont{and} \bibinfo{author}{\bibfnamefont{V.~G.}
  \bibnamefont{{Pal'chikov}}}, \bibinfo{journal}{Phys.~Rev.~Lett.}
  \textbf{\bibinfo{volume}{97}}, \bibinfo{pages}{173601}
  (\bibinfo{year}{2006}).

\bibitem[{\citenamefont{Ovsiannikov et~al.}(2013)\citenamefont{Ovsiannikov,
  Pal'chikov, Taichenachev, Yudin, and Katori}}]{ovsiannikov13a}
\bibinfo{author}{\bibfnamefont{V.~D.} \bibnamefont{Ovsiannikov}},
  \bibinfo{author}{\bibfnamefont{V.~G.} \bibnamefont{Pal'chikov}},
  \bibinfo{author}{\bibfnamefont{A.~V.} \bibnamefont{Taichenachev}},
  \bibinfo{author}{\bibfnamefont{V.~I.} \bibnamefont{Yudin}}, \bibnamefont{and}
  \bibinfo{author}{\bibfnamefont{H.}~\bibnamefont{Katori}},
  \bibinfo{journal}{Phys. Rev. A} \textbf{\bibinfo{volume}{88}},
  \bibinfo{pages}{013405} (\bibinfo{year}{2013}).

\bibitem[{\citenamefont{Yang et~al.}(2016)\citenamefont{Yang, He, Guo, Xu,
  Wang, Sheng, Liu, Wang, Derevianko, and Zhan}}]{zhan16a}
\bibinfo{author}{\bibfnamefont{J.}~\bibnamefont{Yang}},
  \bibinfo{author}{\bibfnamefont{X.}~\bibnamefont{He}},
  \bibinfo{author}{\bibfnamefont{R.}~\bibnamefont{Guo}},
  \bibinfo{author}{\bibfnamefont{P.}~\bibnamefont{Xu}},
  \bibinfo{author}{\bibfnamefont{K.}~\bibnamefont{Wang}},
  \bibinfo{author}{\bibfnamefont{C.}~\bibnamefont{Sheng}},
  \bibinfo{author}{\bibfnamefont{M.}~\bibnamefont{Liu}},
  \bibinfo{author}{\bibfnamefont{J.}~\bibnamefont{Wang}},
  \bibinfo{author}{\bibfnamefont{A.}~\bibnamefont{Derevianko}},
  \bibnamefont{and} \bibinfo{author}{\bibfnamefont{M.}~\bibnamefont{Zhan}},
  \bibinfo{journal}{Phys. Rev. Lett.} \textbf{\bibinfo{volume}{117}},
  \bibinfo{pages}{123201} (\bibinfo{year}{2016}).

\bibitem[{\citenamefont{Sheng et~al.}(2018)\citenamefont{Sheng, He, Xu, Guo,
  Wang, Xiong, Liu, Wang, and Zhan}}]{zhan18a}
\bibinfo{author}{\bibfnamefont{C.}~\bibnamefont{Sheng}},
  \bibinfo{author}{\bibfnamefont{X.}~\bibnamefont{He}},
  \bibinfo{author}{\bibfnamefont{P.}~\bibnamefont{Xu}},
  \bibinfo{author}{\bibfnamefont{R.}~\bibnamefont{Guo}},
  \bibinfo{author}{\bibfnamefont{K.}~\bibnamefont{Wang}},
  \bibinfo{author}{\bibfnamefont{Z.}~\bibnamefont{Xiong}},
  \bibinfo{author}{\bibfnamefont{M.}~\bibnamefont{Liu}},
  \bibinfo{author}{\bibfnamefont{J.}~\bibnamefont{Wang}}, \bibnamefont{and}
  \bibinfo{author}{\bibfnamefont{M.}~\bibnamefont{Zhan}},
  \bibinfo{journal}{Phys. Rev. Lett.} \textbf{\bibinfo{volume}{121}},
  \bibinfo{pages}{240501} (\bibinfo{year}{2018}).

\bibitem[{\citenamefont{Ushijima et~al.}(2018)\citenamefont{Ushijima, Takamoto,
  and Katori}}]{ushijima18a}
\bibinfo{author}{\bibfnamefont{I.}~\bibnamefont{Ushijima}},
  \bibinfo{author}{\bibfnamefont{M.}~\bibnamefont{Takamoto}}, \bibnamefont{and}
  \bibinfo{author}{\bibfnamefont{H.}~\bibnamefont{Katori}},
  \bibinfo{journal}{Phys. Rev. Lett.} \textbf{\bibinfo{volume}{121}},
  \bibinfo{pages}{263202} (\bibinfo{year}{2018}).

\bibitem[{\citenamefont{Kulosa et~al.}(2015)\citenamefont{Kulosa, Fim, Zipfel,
  R\"uhmann, Sauer, Jha, Gibble, Ertmer, Rasel, Safronova et~al.}}]{kulosa15a}
\bibinfo{author}{\bibfnamefont{A.~P.} \bibnamefont{Kulosa}},
  \bibinfo{author}{\bibfnamefont{D.}~\bibnamefont{Fim}},
  \bibinfo{author}{\bibfnamefont{K.~H.} \bibnamefont{Zipfel}},
  \bibinfo{author}{\bibfnamefont{S.}~\bibnamefont{R\"uhmann}},
  \bibinfo{author}{\bibfnamefont{S.}~\bibnamefont{Sauer}},
  \bibinfo{author}{\bibfnamefont{N.}~\bibnamefont{Jha}},
  \bibinfo{author}{\bibfnamefont{K.}~\bibnamefont{Gibble}},
  \bibinfo{author}{\bibfnamefont{W.}~\bibnamefont{Ertmer}},
  \bibinfo{author}{\bibfnamefont{E.~M.} \bibnamefont{Rasel}},
  \bibinfo{author}{\bibfnamefont{M.~S.} \bibnamefont{Safronova}},
  \bibnamefont{et~al.}, \bibinfo{journal}{Phys. Rev. Lett.}
  \textbf{\bibinfo{volume}{115}}, \bibinfo{pages}{240801}
  (\bibinfo{year}{2015}).

\bibitem[{\citenamefont{{Middelmann} et~al.}(2012)\citenamefont{{Middelmann},
  {Falke}, {Lisdat}, and {Sterr}}}]{middelmann12a}
\bibinfo{author}{\bibfnamefont{T.}~\bibnamefont{{Middelmann}}},
  \bibinfo{author}{\bibfnamefont{S.}~\bibnamefont{{Falke}}},
  \bibinfo{author}{\bibfnamefont{C.}~\bibnamefont{{Lisdat}}}, \bibnamefont{and}
  \bibinfo{author}{\bibfnamefont{U.}~\bibnamefont{{Sterr}}},
  \bibinfo{journal}{Phys.~Rev.~Lett.} \textbf{\bibinfo{volume}{109}},
  \bibinfo{eid}{263004} (\bibinfo{year}{2012}).

\bibitem[{\citenamefont{{Sherman} et~al.}(2012)\citenamefont{{Sherman},
  {Lemke}, {Hinkley}, {Pizzocaro}, {Fox}, {Ludlow}, and {Oates}}}]{sherman12a}
\bibinfo{author}{\bibfnamefont{J.~A.} \bibnamefont{{Sherman}}},
  \bibinfo{author}{\bibfnamefont{N.~D.} \bibnamefont{{Lemke}}},
  \bibinfo{author}{\bibfnamefont{N.}~\bibnamefont{{Hinkley}}},
  \bibinfo{author}{\bibfnamefont{M.}~\bibnamefont{{Pizzocaro}}},
  \bibinfo{author}{\bibfnamefont{R.~W.} \bibnamefont{{Fox}}},
  \bibinfo{author}{\bibfnamefont{A.~D.} \bibnamefont{{Ludlow}}},
  \bibnamefont{and} \bibinfo{author}{\bibfnamefont{C.~W.}
  \bibnamefont{{Oates}}}, \bibinfo{journal}{Phys.~Rev.~Lett.}
  \textbf{\bibinfo{volume}{108}}, \bibinfo{eid}{153002} (\bibinfo{year}{2012}).

\bibitem[{\citenamefont{Kramida et~al.}()\citenamefont{Kramida, Ralchenko,
  Reader, and {NIST ASD Team}}}]{nistasd500}
\bibinfo{author}{\bibfnamefont{A.}~\bibnamefont{Kramida}},
  \bibinfo{author}{\bibfnamefont{Y.}~\bibnamefont{Ralchenko}},
  \bibinfo{author}{\bibfnamefont{J.}~\bibnamefont{Reader}}, \bibnamefont{and}
  \bibinfo{author}{\bibnamefont{{NIST ASD Team}}}, \emph{\bibinfo{title}{{NIST
  Atomic Spectra Database (version 5.0.0)}}}.

\bibitem[{\citenamefont{{Porsev} et~al.}(2008)\citenamefont{{Porsev}, {Ludlow},
  {Boyd}, and {Ye}}}]{porsev08a}
\bibinfo{author}{\bibfnamefont{S.~G.} \bibnamefont{{Porsev}}},
  \bibinfo{author}{\bibfnamefont{A.~D.} \bibnamefont{{Ludlow}}},
  \bibinfo{author}{\bibfnamefont{M.~M.} \bibnamefont{{Boyd}}},
  \bibnamefont{and} \bibinfo{author}{\bibfnamefont{J.}~\bibnamefont{{Ye}}},
  \bibinfo{journal}{Phys.~Rev.~A} \textbf{\bibinfo{volume}{78}},
  \bibinfo{pages}{032508} (\bibinfo{year}{2008}).

\bibitem[{\citenamefont{Porsev et~al.}(2017)\citenamefont{Porsev, Safronova,
  and Safronova}}]{porsev17a}
\bibinfo{author}{\bibfnamefont{S.~G.} \bibnamefont{Porsev}},
  \bibinfo{author}{\bibfnamefont{U.~I.} \bibnamefont{Safronova}},
  \bibnamefont{and} \bibinfo{author}{\bibfnamefont{M.~S.}
  \bibnamefont{Safronova}}, \bibinfo{journal}{Phys. Rev. A}
  \textbf{\bibinfo{volume}{96}}, \bibinfo{pages}{012509}
  (\bibinfo{year}{2017}).

\bibitem[{\citenamefont{Ovsiannikov et~al.}(2017)\citenamefont{Ovsiannikov,
  Marmo, Mokhnenko, and Palchikov}}]{ovsiannikov17a}
\bibinfo{author}{\bibfnamefont{V.~D.} \bibnamefont{Ovsiannikov}},
  \bibinfo{author}{\bibfnamefont{S.~I.} \bibnamefont{Marmo}},
  \bibinfo{author}{\bibfnamefont{S.~N.} \bibnamefont{Mokhnenko}},
  \bibnamefont{and} \bibinfo{author}{\bibfnamefont{V.~G.}
  \bibnamefont{Palchikov}}, \bibinfo{journal}{J. Phys.: Conf. Ser.}
  \textbf{\bibinfo{volume}{793}}, \bibinfo{pages}{012020}
  (\bibinfo{year}{2017}).

\bibitem[{\citenamefont{Porsev et~al.}(2018)\citenamefont{Porsev, Safronova,
  Safronova, and Kozlov}}]{porsev18a}
\bibinfo{author}{\bibfnamefont{S.~G.} \bibnamefont{Porsev}},
  \bibinfo{author}{\bibfnamefont{M.~S.} \bibnamefont{Safronova}},
  \bibinfo{author}{\bibfnamefont{U.~I.} \bibnamefont{Safronova}},
  \bibnamefont{and} \bibinfo{author}{\bibfnamefont{M.~G.}
  \bibnamefont{Kozlov}}, \bibinfo{journal}{Phys. Rev. Lett.}
  \textbf{\bibinfo{volume}{120}}, \bibinfo{pages}{063204}
  (\bibinfo{year}{2018}).

\bibitem[{\citenamefont{Tang et~al.}(2014{\natexlab{a}})\citenamefont{Tang, Li,
  and Qiao}}]{tang14a}
\bibinfo{author}{\bibfnamefont{Y.-B.} \bibnamefont{Tang}},
  \bibinfo{author}{\bibfnamefont{C.-B.} \bibnamefont{Li}}, \bibnamefont{and}
  \bibinfo{author}{\bibfnamefont{H.-X.} \bibnamefont{Qiao}},
  \bibinfo{journal}{Chin. Phys. B} \textbf{\bibinfo{volume}{23}},
  \bibinfo{pages}{063101} (\bibinfo{year}{2014}{\natexlab{a}}).

\bibitem[{\citenamefont{{\"{O}}pik}(1967)}]{opik67a}
\bibinfo{author}{\bibfnamefont{U.}~\bibnamefont{{\"{O}}pik}},
  \bibinfo{journal}{Proc.~Phys.~Soc.~London} \textbf{\bibinfo{volume}{92}},
  \bibinfo{pages}{566} (\bibinfo{year}{1967}).

\bibitem[{\citenamefont{Johnson et~al.}(1988)\citenamefont{Johnson, Blundell,
  and Sapirstein}}]{johnson88a}
\bibinfo{author}{\bibfnamefont{W.~R.} \bibnamefont{Johnson}},
  \bibinfo{author}{\bibfnamefont{S.~A.} \bibnamefont{Blundell}},
  \bibnamefont{and}
  \bibinfo{author}{\bibfnamefont{J.}~\bibnamefont{Sapirstein}},
  \bibinfo{journal}{Phys.~Rev.~A} \textbf{\bibinfo{volume}{37}},
  \bibinfo{pages}{307} (\bibinfo{year}{1988}).

\bibitem[{\citenamefont{Wu et~al.}(2019)\citenamefont{Wu, Tang, Shi, and
  Tang}}]{wu19b}
\bibinfo{author}{\bibfnamefont{F.-F.} \bibnamefont{Wu}},
  \bibinfo{author}{\bibfnamefont{Y.-B.} \bibnamefont{Tang}},
  \bibinfo{author}{\bibfnamefont{T.-Y.} \bibnamefont{Shi}}, \bibnamefont{and}
  \bibinfo{author}{\bibfnamefont{L.-Y.} \bibnamefont{Tang}},
  \bibinfo{journal}{Phys. Rev. A} \textbf{\bibinfo{volume}{100}},
  \bibinfo{pages}{042514} (\bibinfo{year}{2019}).

\bibitem[{\citenamefont{{Porsev} et~al.}(2004)\citenamefont{{Porsev},
  {Derevianko}, and {Fortson}}}]{porsev04a}
\bibinfo{author}{\bibfnamefont{S.~G.} \bibnamefont{{Porsev}}},
  \bibinfo{author}{\bibfnamefont{A.}~\bibnamefont{{Derevianko}}},
  \bibnamefont{and} \bibinfo{author}{\bibfnamefont{E.~N.}
  \bibnamefont{{Fortson}}}, \bibinfo{journal}{Phys.~Rev.~A}
  \textbf{\bibinfo{volume}{69}}, \bibinfo{pages}{021403(R)}
  (\bibinfo{year}{2004}).

\bibitem[{\citenamefont{Tang et~al.}(2014{\natexlab{b}})\citenamefont{Tang,
  Yan, Shi, and Babb}}]{tang14b}
\bibinfo{author}{\bibfnamefont{L.-Y.} \bibnamefont{Tang}},
  \bibinfo{author}{\bibfnamefont{Z.-C.} \bibnamefont{Yan}},
  \bibinfo{author}{\bibfnamefont{T.-Y.} \bibnamefont{Shi}}, \bibnamefont{and}
  \bibinfo{author}{\bibfnamefont{J.~F.} \bibnamefont{Babb}},
  \bibinfo{journal}{Phys. Rev. A} \textbf{\bibinfo{volume}{90}},
  \bibinfo{pages}{012524} (\bibinfo{year}{2014}{\natexlab{b}}).

\bibitem[{\citenamefont{Babb}(2015)}]{babb15a}
\bibinfo{author}{\bibfnamefont{J.~F.} \bibnamefont{Babb}},
  \bibinfo{journal}{Phys. Rev. A} \textbf{\bibinfo{volume}{92}},
  \bibinfo{pages}{022712} (\bibinfo{year}{2015}).

\bibitem[{\citenamefont{Tang et~al.}(2013)\citenamefont{Tang, Qiao, Shi, and
  Mitroy}}]{tang13b}
\bibinfo{author}{\bibfnamefont{Y.-B.} \bibnamefont{Tang}},
  \bibinfo{author}{\bibfnamefont{H.-X.} \bibnamefont{Qiao}},
  \bibinfo{author}{\bibfnamefont{T.-Y.} \bibnamefont{Shi}}, \bibnamefont{and}
  \bibinfo{author}{\bibfnamefont{J.}~\bibnamefont{Mitroy}},
  \bibinfo{journal}{Phys. Rev. A} \textbf{\bibinfo{volume}{87}},
  \bibinfo{pages}{042517} (\bibinfo{year}{2013}).

\bibitem[{\citenamefont{Jiang et~al.}(2016)\citenamefont{Jiang, Mitroy, Cheng,
  and Bromley}}]{jiang16a}
\bibinfo{author}{\bibfnamefont{J.}~\bibnamefont{Jiang}},
  \bibinfo{author}{\bibfnamefont{J.}~\bibnamefont{Mitroy}},
  \bibinfo{author}{\bibfnamefont{Y.}~\bibnamefont{Cheng}}, \bibnamefont{and}
  \bibinfo{author}{\bibfnamefont{M.~W.~J.} \bibnamefont{Bromley}},
  \bibinfo{journal}{Phys. Rev. A} \textbf{\bibinfo{volume}{94}},
  \bibinfo{pages}{062514} (\bibinfo{year}{2016}).

\bibitem[{\citenamefont{Jiang et~al.}(2013)\citenamefont{Jiang, Tang, and
  Mitroy}}]{jiang13a}
\bibinfo{author}{\bibfnamefont{J.}~\bibnamefont{Jiang}},
  \bibinfo{author}{\bibfnamefont{L.~Y.} \bibnamefont{Tang}}, \bibnamefont{and}
  \bibinfo{author}{\bibfnamefont{J.}~\bibnamefont{Mitroy}},
  \bibinfo{journal}{Phys. Rev. A} \textbf{\bibinfo{volume}{87}},
  \bibinfo{pages}{032518} (\bibinfo{year}{2013}).

\end{thebibliography}

\end{document}